# A plasmonic ellipse resonator possessing hybrid modes for ultracompact chipscale application

Zhaojian Zhang [1], Junbo Yang [2] *, Heng Xu [1], Siyu Xu [1], Yunxin Han [2], Xin He [2], Jingjing Zhang [1], Jie Huang [1], Dingbo Chen [1] and Wanlin Xie [1]

[1] College of Liberal Arts and Sciences, National University of Defense Technology, Changsha 410073, China

[2] Center of Material Science, National University of Defense Technology, Changsha 410073, China

* Correspondence: yangjunbo@nudt.edu.cn; 376824388@qq.com

**Abstract:** We propose a plasmonic ellipse resonator possessing hybrid modes based on metal-insulator-metal (MIM) waveguide system. Specially, this nanocavity has hybrid characteristic of rectangle and disk resonator, therefore supporting both Fabry-Perot modes (FPMs) and whispering-gallery modes (WGMs). Besides, by changing the length of major and minor radius of the ellipse, the resonant wavelengths of FPMs and WGMs can be independently tuned and close to each other, thus constructing a plasmon-induced transparency (PIT) - like spectrum profile. Benefitting from this, a dual-band slow light is achieved with one single resonator. Furthermore, this component can also act as a multi-band color filter and refractive index sensor. We believe such multi-mode resonator with ultra-small footprint will play an important role in more compact on-chip optical circuits in the future.

## 1. Introduction

Arising from the interaction between free electrons of metal and electromagnetic waves, surface plasmon polariton (SPP) is the collective electromagnetic oscillation supported on the interface of metal and insulator. SPP has drawn more attention due to the capability to manipulate light beyond the diffraction limit, which is promising for photonic devices within subwavelength scale [1]. Such plasmonic components will play an important role in more compact photonic systems, and various applications have been explored such as sensors, lasers and modulators [2-4].

Recently, as an ideal platform to realize highly integrated optical circuits, plasmonic metal-insulator-metal (MIM) waveguide system has been studied widely, because it possesses advantages of deep confinement of light, considerate propagation length as well as easy fabrication [5]. Based on such system, different kinds of on-chip plasmonic devices have been realized. Utilizing one or several coupled resonators, single- or multi-band filters and demultiplexers are achieved [6-8]. Via the coupling between multiple resonators, Fano resonance is introduced and applied for sensing [9-11]. As a special case of Fano resonance, plasmon-induced transparency (PIT) has also been focused, which can lead to a slow light effect in the transparent window [12-14]. However, as the improvement of integration, devices with smaller footprint and simpler framework will be needed, and multiple-resonator system cannot meet such requirement.

Here, we propose a plasmonic ellipse nanoresonator possessing hybrid modes based on MIM waveguide system. Such resonator possesses characteristics of both rectangle and disk cavity, thus supporting Fabry-Perot modes (FPMs) as well as whispering-gallery modes (WGMs) simultaneously. Besides, via changing the length of major and minor radius of the ellipse, the resonant wavelengths of FPMs and WGMs can be independently tuned and close to each other, leading to a PIT-like spectrum profile. Thanks to this, a dual-band slow light effect is achieved in this one single resonator. Furthermore, such multi-mode resonator can also act as a multi-band color filter and sensor. This device possesses multiple functions, ultracompact configuration as well as easy integration, and will find potential applications on future on-chip optical circuits.

## 2. Structures and methods

The proposed structure is depicted in Fig. 1, consisting of an input waveguide and a side-coupled ellipse resonator. The geometric parameters are as follows: major radius $R$= 280 nm, minor radius $r$= 140 nm, the gap $g$= 15 nm, waveguide width $w$= 50 nm, and the length of the device $D$= 1000 nm. Here, the grey area is silver (Ag), and white area is air. The permittivity of Ag is characterized by the Drude model [15]:

$$\varepsilon_m = \varepsilon_\infty - \frac{\omega_p^2}{\omega(\omega+i\gamma)} \quad (1)$$

Here, $\varepsilon_\infty$ is permittivity at infinite frequency, $\omega_p$ is bulk plasma frequency, $\gamma$ is electron oscillation damping frequency, $\omega$ is angular frequency of incident waves. For Ag, the parameters are as follows: $\varepsilon_\infty$=3.7, $\omega_p$= 1.38×10$^{16}$ Hz, and $\gamma$= 2.73×10$^{13}$ Hz.

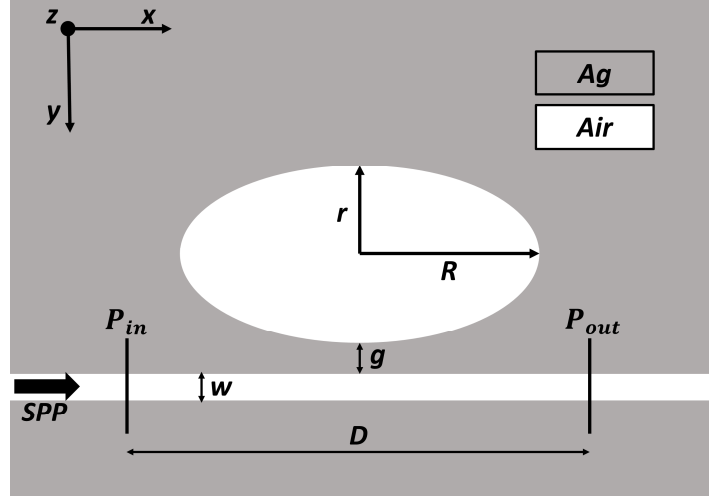

Fig. 1 The 2D schematic of the structure.

In the MIM waveguide, only transverse-magnetic (TM) mode can be supported [16]. Furthermore, there is only fundamental TM mode because the waveguide width is much smaller than the input wavelength. The dispersion of fundamental mode is described as follows [17]:

$$\frac{\varepsilon_i p}{\varepsilon_m k} = \frac{1-e^{kw}}{1+e^{kw}}$$

$$k = k_0\sqrt{(\frac{\beta_{spp}}{k_0})^2 - \varepsilon_i}, p = k_0\sqrt{(\frac{\beta_{spp}}{k_0})^2 - \varepsilon_m} \quad (2)$$

$$\beta_{spp} = n_{eff}k_0 = n_{eff}\frac{2\pi}{\lambda}$$

Here, $\lambda$ is the incident wavelength in vacuum, $\varepsilon_i$ and $\varepsilon_m$ are dielectric and metal constant,

respectively. $\beta_{spp}$ is the propagation constant of SPP, $n_{eff}$ is the effective refractive index of MIM waveguide, and $k_0 = 2\pi/\lambda$ is the wavenumber. Finite-Difference Time-Domain (FDTD) method with mesh size 3 nm is adapted to do the simulation, the boundary condition is set as perfectly matched layers (PML). As shown in Fig. 1, two power monitors are placed at $P_{in}$ and $P_{out}$, so the transmission is defined as $T = P_{out}/P_{in}$.

## 3. Results and discussion

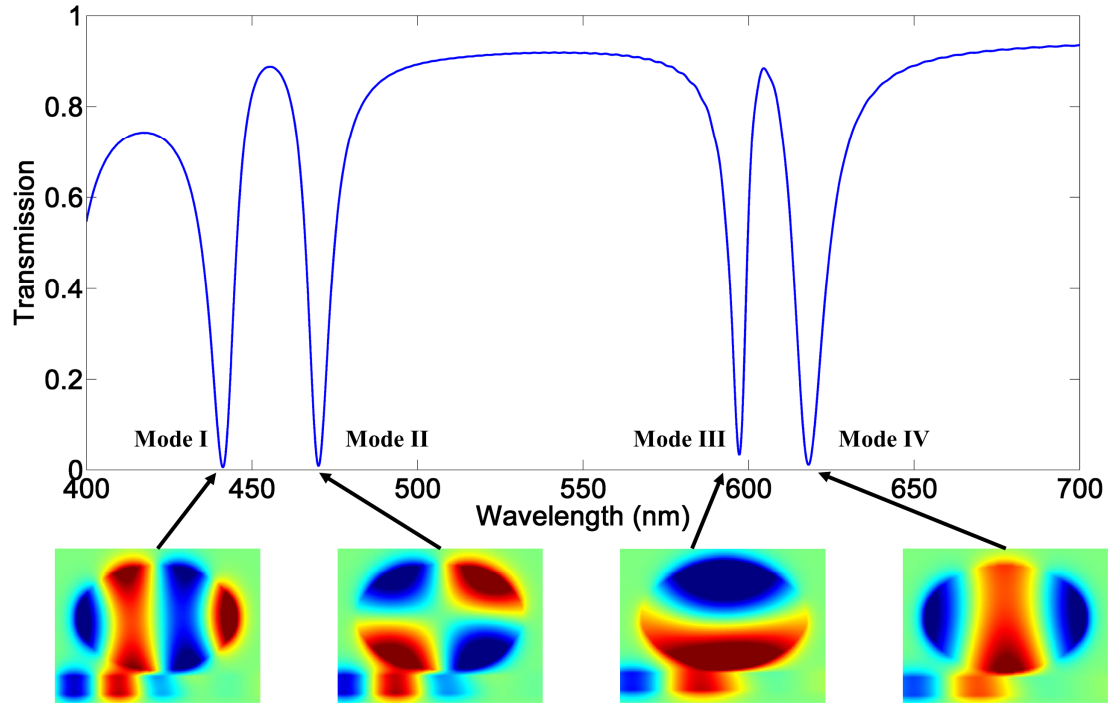

Fig. 2 The transmission spectrum of the structure and $H_z$ distributions of the four modes.

The transmission spectrum of this structure is given in Fig. 2, showing that such ellipse resonator possesses four modes in the visible light band from 400 to 700 nm. The $H_z$ field distribution of each mode is depicted in the bottom of Fig. 2, from which we can see that, mode I at 441.17 nm and mode IV at 618.08 nm possess the field characteristic of FPMs, meanwhile mode II at 470.03 nm and mode III at 597.20 nm have the field patterns of WGMs. Normally, a rectangle resonator can support FPMs [18], and a disk resonator can have WGMs [19]. Therefore, such ellipse resonator possesses a hybrid characteristic of both rectangle and disk resonator.

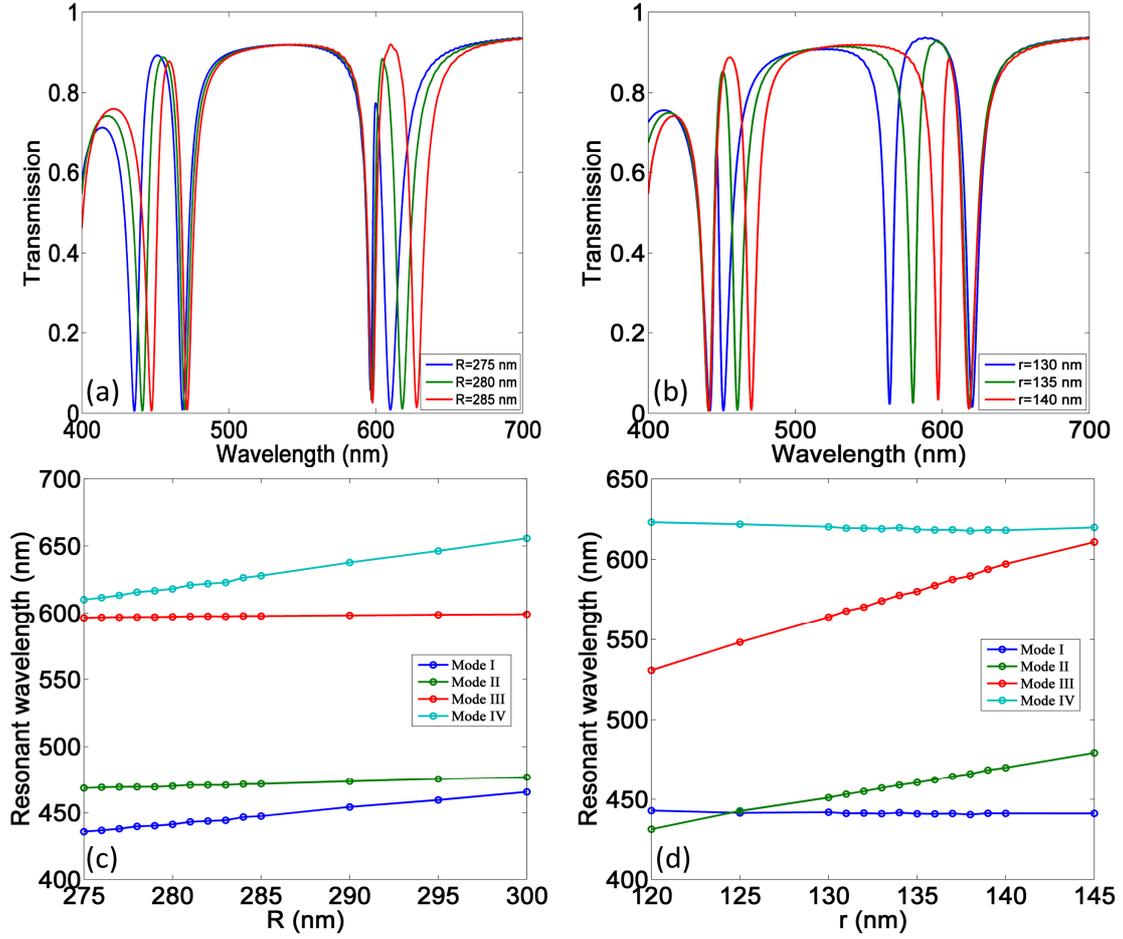

Fig.3 (a) The spectra under different major radiuses. (b) The spectra under different minor radiuses. (c)The relation between resonant wavelength and major radius. (d) The relation between resonant wavelength and minor radius.

Next, the influence of geometric parameters on the resonant wavelengths is investigated. At first, the minor radius of the ellipse is maintained as 140 nm, and the major radius is changed from 275 nm to 300 nm. The transmission spectra under different major radiuses are shown in Fig. 3(a), and the relations between resonant wavelengths of the four modes and major radiuses are presented in Fig. 3(c). Apparently, the resonant wavelengths of mode I and IV, which are FPMs, are sensitive to the major radius and the relation is linear, while the mode II and III, which are WGMs, are insensitive. Then, the major radius is kept as 280 nm, and the minor radius varies from 130 to 140 nm. The corresponding different spectra as well as the relation between resonant wavelength and minor radius are shown in Fig. 3(b) and (d). In this case, however, mode II and III are sensitive to the minor radius while mode I and IV are insensitive.

This is because FPM is formed by standing waves, which are reflected back and forth between the left and right end of the ellipse resonator. Therefore, the resonant wavelength of FPM is decided by follows [20]:

$$m\lambda_r = 2n_{eff}R_{eff}, \quad m = 1, 2......$$  (3)

Where $\lambda_r$ is the resonant wavelength of the FPM, $n_{eff}$ is the effective refractive index inside the ellipse, $R_{eff}$ is the effective major radius including the penetration depths, dominated by major radius. $m$ is the mode order number, which is an integer. From this equation, the resonant wavelength of the same order FPM has a linear relation with the major radius, which accords with the result shown in Fig. 3 (c). From the mode patterns shown in Fig. 2, we can see that mode I is the third order FPM and mode IV is the second order FPM. However, for the WGMs, the resonant condition is as follows [19]:

$$n\lambda_r = n_{eff} L_{eff}, n = 1, 2... \quad (4)$$

Here, $L_{eff}$ is the effective circumference of ellipse considering the penetration depths, which is dominated by major and minor radius. $n$ is the mode order number, which is an integer. Since minor radius contributes more to the circumference than major radius, WGMs are much more sensitive to the minor radius. From the mode patterns, mode II is the second order WGM and mode III is the first order WGM. Therefore, FPMs and WGMs in this ellipse resonator could be tuned almost independently by changing the major or minor radius. Since the four modes almost cover the whole visible light band, such resonator can act as a multi-band color filter.

## 4. Applications on slow light and sensing

Benefitting from the independently tunable property, FPM and WGM can be arranged to be close, constructing a PIT-like transmission spectrum profile [12]. Such profile indicates an intense dispersion within the transparent window, which will lead to slow light effect. Here, the major and minor radius is set as 280 nm and 140 nm respectively to introduce such analogue of PIT, the slow light can be assessed by the optical delay time $\tau_g$ [12]:

$$\tau_g = \frac{d\psi(\omega)}{d\omega} \quad (5)$$

Here, $\psi(\omega)$ is the transmission phase shift from the light source to the monitor. The phase shift and delay time are shown in Fig. 4(a-b), confirming that there is slow light effect in the two transparent windows, the delay time can be above 0.05 ps at each window. Compared with other multi-band slow light devices [12, 21-22], this work possesses considerate delay time, meanwhile owning a much smaller and simpler framework.

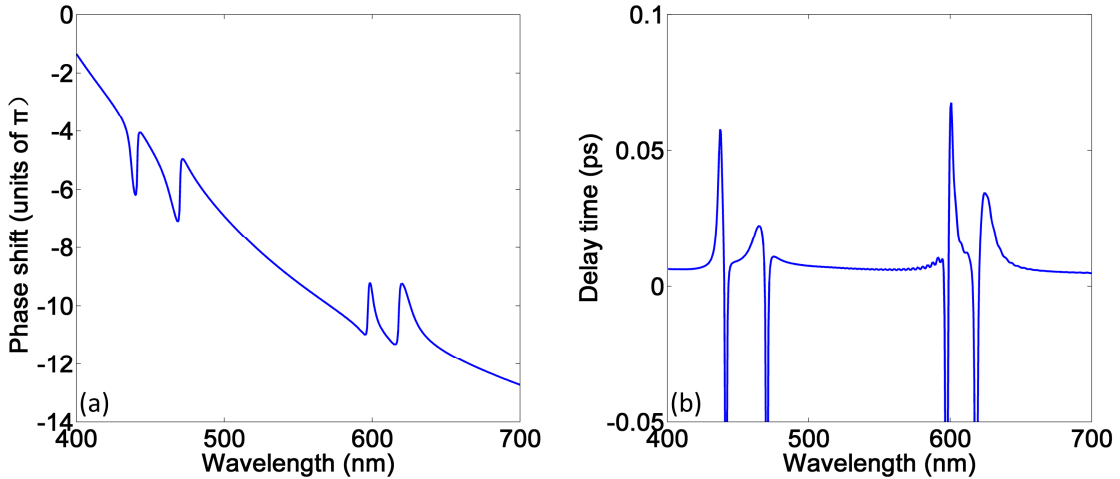

Fig. 4 (a) The phase shift at different wavelengths. (b) The delay time at different wavelengths.

Furthermore, the four modes almost cover the whole visible light band, therefore providing multi-band sensing ability over a broadband range. The sensing performance is assessed by two factors, sensitivity (S) and figure of merit (FOM) [23]:

$$S = \frac{\Delta\lambda}{\Delta n}$$
$$\text{FOM} = \frac{S}{\text{FWHM}} \quad (6)$$

Here, S is the wavelength shift induced by unit change of surrounding refractive index (SRI), FWHM is the width at half maximum, FOM represents the optical resolution of the sensor. The transmission spectra of the four modes under different SRI are shown in Fig. 5(a), showing a red shift with increasing SRI. Fig. 5(b) indicates that resonant wavelengths of the four modes have linear relations with SRI. The sensitivity, FWHM and FOM of four modes are given in table. 1.

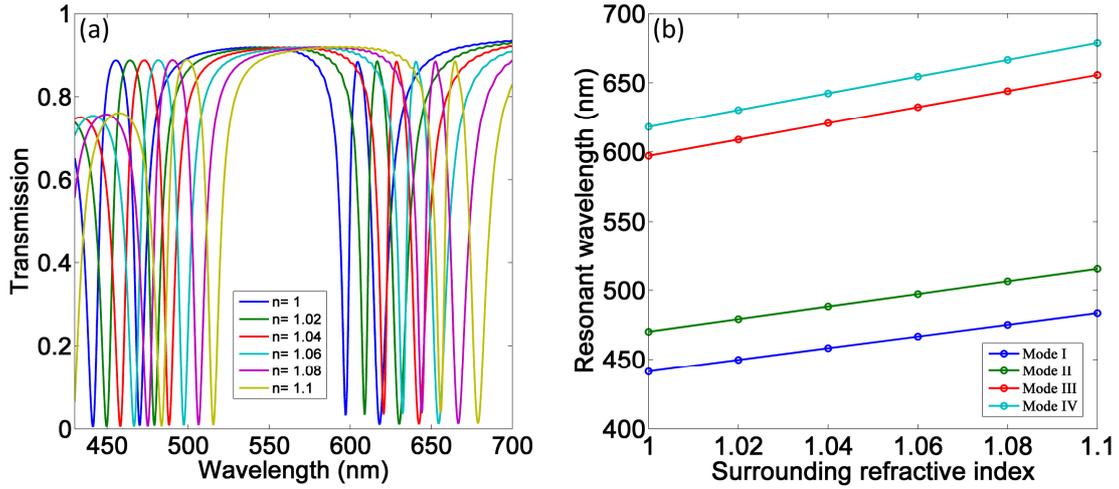

Fig. 5 (a) The transmission spectra under different SRI. (b) The relation between resonant wavelengths and SRI.

Table. 1 The sensitivity, FWHM and FOM of the four modes.

| Mode | I | II | III | IV |
| --- | --- | --- | --- | --- |
| S (nm/RIU) | 424 | 454 | 586 | 608 |
| FWHM (nm) | 9.48 | 7.92 | 5.58 | 10.56 |
| FOM (/RIU) | 44.73 | 57.32 | 105.02 | 57.58 |

**5. Conclusion**

In summary, we propose an ellipse resonator in the MIM waveguide system. Such resonator possesses hybrid modes, including FPMs and WGMs, meantime has ultracompact configuration and small footprint. Benefiting from the independently tunability of different modes, dual-band slow light is achieved. Furthermore, such resonator can act as a multi-band color filter and sensor in the visible light band. This work can find potential applications on highly integrated optical circuits.

**Acknowledgments**

This work is supported by the National Natural Science Foundation of China (61671455, 61805278), the Foundation of NUDT (ZK17-03-01), the Program for New Century Excellent Talents in University (NCET-12-0142), and the China Postdoctoral Science Foundation (2018M633704).

with high figure of merit based on concentric-rings resonator. Sensors, 18(1), 116.